\documentclass[a4paper,twocolumn,amsmath,pra,showpacs,amssymb]{revtex4-1} 
\usepackage[T1]{fontenc}
\usepackage[latin9]{inputenc}
\usepackage{times}
\usepackage{color}
\usepackage{xspace}
\usepackage{amssymb,amsmath}
\usepackage{amsbsy}
\usepackage{graphicx}
\usepackage{multirow}
\usepackage{rotating}
\usepackage{microtype}
\usepackage{tabularx}
\usepackage{epstopdf}
\usepackage{float}
\usepackage{mathrsfs}
\usepackage{hyperref}
\usepackage[normalem]{ulem}

\usepackage{calligra}
\DeclareMathAlphabet{\mathcalligra}{T1}{calligra}{m}{n}
\DeclareFontShape{T1}{calligra}{m}{n}{<->s*[2.2]callig15}{}

\def\be{\begin{equation}}
\def\ee{\end{equation}}
\def\e#1{\label{#1}\end{equation}}
\def\bea{\begin{eqnarray}}
\def\eea{\end{eqnarray}}
\def\ea#1{\label{#1}\end{eqnarray}}

\def\bem#1{\begin{mathletters}\label{#1}}
\def\eml{\end{mathletters}}

\def\ket#1{{|#1\rangle}}
\def\bra#1{{\langle#1|}}

\def\4#1{{\boldsymbol{#1}}}
\def\8#1{{\widetilde{#1}}}
\def\bse{\begin{subequations}}
\def\ese{\end{subequations}}

\def\Rb87{$^{87}\text{Rb}$}
\def\0{\ket{0}}
\def\1{\ket{1}}

\begin{document}
\title{Deterministic entanglement of Rydberg ensembles by engineered dissipation}
\author{D. D. Bhaktavatsala Rao}
\author{Klaus M{\o}lmer}
\affiliation{%
Department of Physics and Astronomy, Aarhus University, Ny Munkegade 120, DK-8000, Aarhus C, Denmark. \\
}%
\date{\today}
\begin{abstract}
We propose a scheme that employs dissipation to deterministically generate entanglement in an ensemble of strongly interacting Rydberg atoms.
With a combination of microwave driving between different Rydberg levels and a resonant laser coupling to a short lived atomic state, the ensemble
can be driven towards a dark steady state that entangles all atoms. The long-range resonant dipole-dipole interaction between different Rydberg states
extends the entanglement beyond the van der Walls interaction range with perspectives for entangling large and distant ensembles.
 \end{abstract}
 \pacs{03.67.Lx, 32.80.Ee, 32.80.Rm, 42.50.-p}
\maketitle
The strong blockade interaction between Rydberg excited atoms opens  many possibilities to explore neutral atoms for quantum computing and for the study of a variety of
complex many-body and light-matter problems \cite{revmod}. After the first proposal by Jaksch et al.\cite{jaks} to use Rydberg blockade to implement a fast two-qubit
controlled-NOT gate there has been a variety of schemes for fast quantum gates with atomic ensembles \cite{luk, saff1,saff2,zheng}.

 Dissipation has become an alternative component in the creation of complex entangled states \cite{kraus, diehl, davido} and in the implementation of quantum computing \cite{peter, cirac}. The remarkable feature of dissipative approaches is their resilience to errors that occur with  imperfect state initialization, to fluctuations in the driving field strengths and to dependencies on the system size - errors which are devastating for unitary approaches to create the same entangled state \cite{unitary}. Dissipative approaches have been proposed to create entanglement of pairs of qubits \cite{sorensen} and the robustness of dissipatively driven entanglement has been verified in ion traps \cite{wineland} and superconducting circuits \cite{sc}, and in the collective spin degrees of freedom of large atomic ensembles \cite{polzik}.

 Also, proposals have been made to combine dissipation and Rydberg blockade to entangle a pair of neutral atom qubits \cite{durga, saffman}. {\it E.g.}, in \cite{durga} two qubits have a unique "singlet state", $\ket{01}-\ket{10}$, which is invariant under common rotations applied to both qubits, and which makes it the steady state under combined qubit rotations and Rydberg excitation. A similar situation does  not occur for many qubits where dissipation for which the singlet states are dark states is much harder to engineer and where the singlet space is
 degenerate (for even $N>2$). For larger atom numbers, Rydberg blockade and dissipation have been applied to prepare lattice systems with steady state spatial correlations
 \cite{igor, weimer, davidf, lee, phol}. Recent work \cite{choi} has proposed to use the Rydberg blockade interaction to mediate an interaction between atoms in a lattice systems, and by engineered dissipation of atoms on the edge of the lattice drive the rest of the ensemble into an entangled state.

\begin{figure}
\includegraphics[width=90mm]{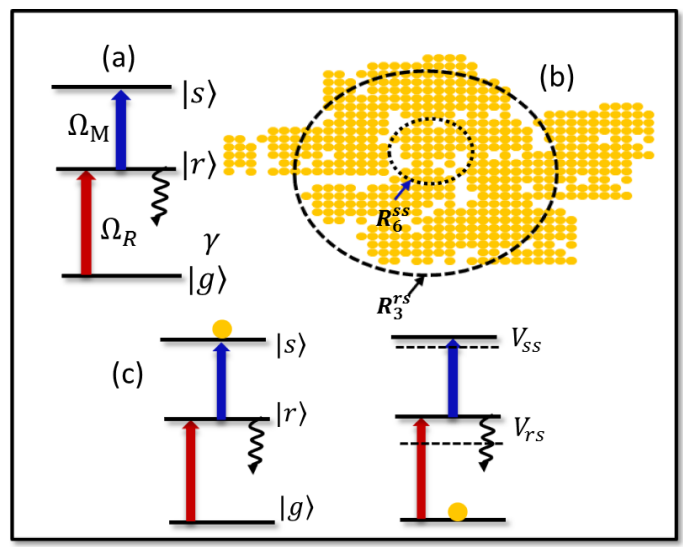}
\label{setup1}
\vspace{-10mm}
\caption{(Color online) (a) Energy-level diagram  for a single atom. The ground state $\ket{g}$ is coupled by the effective
Rabi frequency $\Omega_R$ to the Rydberg state $\ket{r}$, which is further coupled by a microwave field with Rabi frequency $\Omega_M$ to
another Rydberg level $\ket{s}$. The decay rate $\gamma$ of the state $\ket{r}$ is engineered by resonant coupling to a short-lived,
optically excited state $\ket{e}$ (not shown). (b) Schematic representation of a frozen Rydberg ensemble showing the blockade radius $R_6^{ss}$ for the Van der Waals interaction between atoms occupying the same Rydberg state $\ket{s}$
and the blockade radius $R_3^{rs}$ for the resonant dipolar coupling between atoms occupying different Rydberg states $\ket{r}$ and $\ket{s}$.
(c) Schematic representation of how the Rydberg excitation of one atom to the level $\ket{s}$ shifts the energy levels of states $\ket{r}$ and $\ket{s}$ in a neighboring atom.}
\end{figure}

In this Letter we propose a strategy to deterministically prepare an ensemble of atoms in a dark, entangled state. The proposal exploits a combination
of Rydberg blockade, destructive interference in three-level atoms, and engineered dissipation through coupling of long-lived Rydberg states to a short lived optically excited states.

\begin{figure}
\includegraphics[width=80mm]{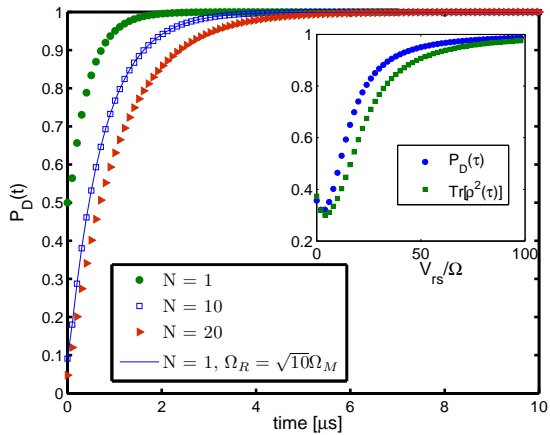}
\label{setup1}
\caption{(Color online)The population $P_D$ of the dark state (3) is plotted as a function of time for various ensemble sizes $N$ initialized in the ground state $\ket{G}$. We assume
a perfect blockade between the atoms occupying any two Rydberg levels i.e., $V_{rr} = V_{ss} = V_{rs} = \infty$ and in our numerical calculations
we have taken field strengths $\Omega_R/2\pi = \Omega_M/2\pi = 1$ MHz and an engineered decay rate $\gamma/2\pi =2$ MHz  of the Rydberg level $\ket{r}$. Also shown in the same figure
is the case of a single atom driven by fields $\Omega_R/2\pi=\sqrt{10}$ MHz, $\Omega_M/2\pi = 1$ MHz. In the inset we show the
dark-state population $P_D(\tau)$ and the purity, $Tr[\rho^2(\tau)]$, of the state as a function of $V_{rs}$ for $N=4$ atoms at a fixed time $\tau = 5\mu$s.}
\end{figure}

Fig. 1(a) shows the atomic level structure. Each atom has a ground state $\ket{g}$ which is coupled with a Rabi frequency $\Omega_R$ to a Rydberg state $\ket{r}$ by a single photon \cite{sandia}
or a two-photon process via a highly detuned intermediate level. The Rydberg level $\ket{r}$ is coupled to an adjacent Rydberg level $\ket{s}$ by a microwave field with Rabi frequency
$\Omega_M$. The Rydberg states are long lived, and we engineer a fast decay channel by coupling $\ket{r}$ resonantly to a short-lived state $\ket{e}$, which decays rapidly
back to the ground state. The resulting effective decay rate, $\gamma$ for the Rydberg level $\ket{r}$ is a function of the $\ket{r}\leftrightarrow\ket{e}$ coupling strength and the decay rate of $\ket{e}$.
In the presence of decay $\gamma$ from the level $\ket{r}$, the steady state solution for a single atom is  $\ket{\psi^{(1)}_D} = \frac{1}{\Omega_1}[\Omega_M\ket{g}-\Omega_R\ket{s}]$,
where $\Omega_1=\sqrt{\Omega_M^2 + \Omega_R^2}$. This state is called a dark state as it has no unstable atomic state components and hence emits no radiation.

To understand the dynamics of an atomic ensemble subject to the interactions mentioned above, let us now consider the two-atom case, illustrated in Fig. 1(c).
When atoms occupy the same Rydberg level $\ket{r(s)}$, they experience the van der Waals interaction, $V_{rr(ss)}\propto 1/r^6$, while if they occupy different states, they
experience the resonant dipole-dipole interaction (RDDI), $V_{rs}\propto 1/r^3$.

If two such atoms are in close proximity the strong blockade interaction $V_{ss}$ does not allow double occupancy of the state $\ket{s}$. Note that $V_{sr}$ has a similar effect:
when one atom is in the state $\ket{s}$ the ground state coupling of the other atom to its $\ket{r}$ state is detuned from resonance due to the strong resonant dipole coupling which
leads to two-atom eigenstates $(\ket{sr}\pm\ket{rs})/\sqrt{2}$ with perturbed energy levels, $\pm V_{rs}$.
This suppresses the evolution of the second atom if it is initialized in the ground state. If the first atom occupies the dark state with population in both $\ket{g}$ and
$\ket{s}$, its $\ket{g}$ component allows excitation of the other atom, and two atoms driven by the same laser and microwave fields have a unique dark state
$\ket{\psi^{(2)}_D}=\frac{1}{\Omega_2}[\Omega_M\ket{gg}-\Omega_R\ket{gs}-\Omega_R\ket{sg}]$, where $\Omega_2=\sqrt{\Omega_M^2 + 2\Omega_R^2}$. The mechanism behind the
convergence of the system into this two-atom dark state differs from other proposals using three-level atoms, where the unstable optically excited state is used as intermediate
state  \cite{davidmolm}. In particular, even if $V_{ss}$ is negligible, our use of an intermediate state which mediates a strong atom-atom interaction, enables the formation of coherent,
multi-atom entangled states.

The Hamiltonian describing an ensemble of atoms subject to the interactions described above can be written as
\bea
 H &=& \sum_k H_k +\sum_{i\ne j}V^{ij}_{rr}\ket{r_ir_j}\bra{r_ir_j} + V^{ij}_{ss}\ket{s_is_j}\bra{s_is_j}  \\ \nonumber
&&   + V^{ij}_{rs}(\ket{r_is_j}\bra{s_ir_j}+ h.c).
 \eea
where the single atom interaction with the laser and microwave fields are given in the corresponding rotating frame by the Hamiltonian $H_k = \left(\Omega_R\ket{g_k}\bra{r_k} + \Omega_M\ket{r_k}\bra{s_k} + h.c\right)$.
In the limit of strong blockade i.e., all interaction strengths $V >> \Omega_{R(M)}$, the dynamics is restricted to states with at most a single Rydberg excited atom. {\it I.e.,} the system will not explore the full $3^N$
dimensional Hilbert space, as only the subspace spanned by $\ket{G} = \ket{gg\cdots g_N}$, $\ket{R_k} = \ket{gg\cdots r_k\cdots g_N}$ and $\ket{S_k} = \ket{gg\cdots s_k\cdots g_N}$ is accessible from the initial ground state $\ket{G}$. On this $2N+1$ dimensional space,  the effective Hamiltonian
\be
H_{eff} = \sum_k\left(\Omega_R\ket{G}\bra{R_k} + \Omega_M\ket{R_k}\bra{S_k} + h.c\right)
\ee
applies, and we identify the unique state with vanishing eigenvalue
\be
\ket{\psi^{(N)}_D} = \frac{1}{\Omega_N}\left[\Omega_M\ket{G}-\sqrt{N}\Omega_R\ket{S}\right],
\ee
where $\Omega_N = \sqrt{\Omega^2_M+N\Omega^2_R}$ and the collective state $\ket{S} \equiv \frac{1}{\sqrt{N}}\sum_kS_k$.

\begin{figure}
\includegraphics[width=80mm]{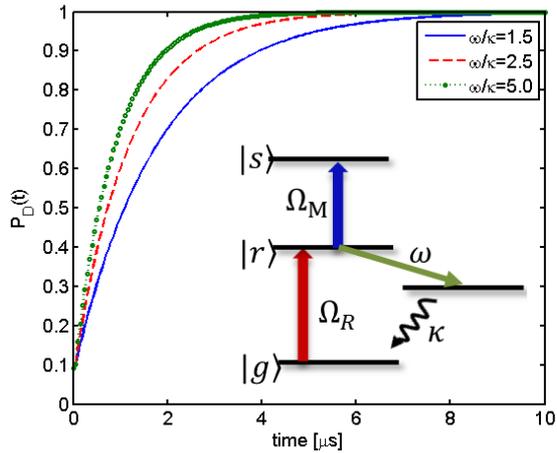}
\label{setup1}
\caption{(Color online) The population $P_D$ of the dark state of an ensemble with $N=10$ atoms is plotted as a function of time for different values of the coupling frequency $\omega$ between the Rydberg state $\ket{r}$ and the optically excited state $\ket{e}$ (see inset).
We have assumed the decay rate $\kappa/2\pi = 5$MHz corresponding to the optically excited state $|5P_{3/2}\rangle$ in Rb and the Rabi and microwave field strengths are taken to be
$\Omega_R/2\pi = \Omega_M/2\pi = 1$MHz.}
\end{figure}

It has been shown \cite{kraus} that one can identify a set of Lindblad dissipation operators that will turn a given  eigenstate of a Hamiltonian into the unique dark, steady state of the dissipative dynamics of the system.
Though generation of such a set is in general a non-trivial task, our case offers a straightforward solution, since
$\ket{\psi^{(N)}_D}$ is the only eigenstate of $H_{eff}$ without components with the short-lived Rydberg excited levels $\ket{R_k}$. The state $\ket{\psi_D}$ is thus the asymptotic steady state reached by the system, as we confirm by solving the master equation
\be
\partial_t \rho = -i[\mathcal{H}_{eff},\rho] + \sum_{j,k} \mathcal{C}^\dagger_k\rho\mathcal{C}_k,
\ee
where $\mathcal{C}_k = \sqrt{\gamma_k}\ket{G}\bra{R_k}$ are Lindblad operators, which describe the decay of the $k^{th}$ atom.
The decay rates $\gamma_k$ do not need to be identical, and even if only one of the atoms decays due to the coupling of the states $\ket{r}$ and $\ket{e}$,
the symmetric dark state $\ket{\psi^{(N)}_D}$ is reached asymptotically (but more slowly than if all atoms have unstable $\ket{r}$ states).
In Fig. 2 we have plotted the dark state population $P_D$ as a function of time for various $N$. We first note the convergence to unit population
of the dark state for all atom numbers, and we observe that as $N$ increases the convergence rate decreases. We also plot in the same figure the dark state population of a single atom driven by
a stronger field $\Omega_R = \sqrt{10}\Omega_M$. This curve coincides with the curve corresponding to $N=10$ and $\Omega_M = \Omega_R$ in accord with the invariance of the dark state (3) under transformations that leave $\sqrt{N}\Omega_R/\Omega_M$ unchanged.

In the above analysis we assumed a perfect blockade preventing the occupancy of any doubly Rydberg excited states, $\ket{rr}, \ket{ss}, \ket{sr(rs)}$.
Though it may appear that the perfect blockade between the states $\ket{rr(ss)}$ is trivially responsible for the formation of the dark state, the role of RDDI in blocking the occupation of states $\ket{rs(sr)}$ is emphasized in the inset of Fig. 2 for $N=4$ atoms. Clearly, when $V_{rs}=0$ both the fidelity and the purity are reduced even though  $V_{rr(ss)} = \infty$. This indicates the absence of a unique dark state in the system, and the steady state solution to the master equation is a mixed state. With increasing $V_{rs}$ the system regains its dark state feature and entanglement. As we shall show below, even when $V_{ss}, ~V_{rr} \sim 0$, a large RDDI suffices to drive the system towards the desired dark state.

It is a key requirement for the protocol that the life time of the state $\ket{r}$ is much shorter than the life time of $\ket{s}$ state. This can be achieved by coupling the state $\ket{r}$ resonantly with Rabi frequency $\omega$ to an optically excited state $\ket{e}$ with lifetime $\kappa$.  In Fig. 3 we show the evolution of the system with $N=10$ atoms with different values of the coupling and decay parameters and corresponding variation in the approach towards the dark steady state. The effective decay rate of the state $\ket{r}$ for the three cases shown in the figure are $\gamma \approx \kappa/5$ (blue solid-line), $\gamma \approx 2\kappa/5$ (red dashed-line) and $\gamma \approx \kappa/2$ (green circles) with $\kappa/2\pi = 5$MHz.

Even though the dark state is a superposition of the ground state $\ket{G}$ and collective state $\ket{S}$, it is robust against dephasing and perturbations of the Rydberg states caused, e.g., by magnetic field fluctuations. We understand this as a consequence of dissipation which may be compared to a continuous measurement process that monitors if the dark state character is maintained. The absence of emission from the $\ket{e}$ state causes suppression of small errors and restoration of the system in dark state, while an emission event is accompanied by a quantum jump into the ground state from where the system evolves back towards the dark state on the time scale $T_f$ shown in Figs 2, 3. The steady state of the system thus suffers a loss of fidelity $\sim (\gamma_d+\gamma_s)T_f$ in presence of such errors with decoherence rate $\gamma_d$ and decay rate $\gamma_s$.

Consider the Rydberg states $\ket{r}=\ket{ns}$ and $\ket{p} = \ket{np}$ with  $n = 70$ in Rubidium. The dipolar couplings $V_{rr}, ~V_{ss}, ~V_{rs}$
between two atoms separated by a distance of $3\mu$m are approximately, $2\pi \times \lbrace 190, 400, 140 \rbrace$MHz,
and if $\Omega_R = \Omega_M = 1$MHz, atoms in an ensemble of this size experience perfect blockade preventing any two atoms to be simultaneously excited to Rydberg states. Using these parameters and considering the typical dephasing and decay rates, $\gamma_d/2\pi = 10$kHz, $\gamma_{s,r}/2\pi = 5$kHz, of the state $\ket{S}$  and an engineered decay rate of the states $\ket{R_k}$ caused by coupling $\omega/2\pi = 24$ MHz to the optically excited state $6P_{1/2}$ with a decay rate $\kappa/2\pi = 6$MHz, we obtain high fidelity dark states with ensembles up to $N = 20$ atoms. For 20 atoms, the dark state (3) is dominated by the entangled state $\ket{S}$, also known as the W-state \cite{bennett}, and with the parameters listed, our simulations yield a population in this state after 10 $\mu$s of $0.914$.

If we instead choose $\Omega_M = \sqrt{20}$MHz, we obtain a dark state with equal weights $\frac{1}{\sqrt{2}}[\ket{G}-\ket{S}]$, and this state has a high fidelity of $\sim 0.988$ and for preparation times $t > 10 \mu$s. Such states can be a starting point for preparation of a Schr\"odinger cat state as shown in \cite{klaus1}. A steady state superposition of a ground and collectively excited state also offers the possibility to release, by a laser pulse on the $\ket{s}\rightarrow \ket{e}$ transition, a phase matched and hence directional photonic qubit state \cite{marksp, klaussp}.

Note that while the time of formation of the dark state decreases with $N$, the population increases monotonically
with time towards the steady state value. Hence, any ensemble of size $N \le 20$ will reach steady state as early as the ensemble with $N=20$. Thus experiments with imprecise knowledge of $N$ or a distribution of ensemble sizes may be carried out.

We now turn to the situation where $V_{rr(ss)} < \Omega_{R,M}$ while $V_{rs} > \Omega_{R,M}$.
Excitation of a pair of atoms to Rydberg states $\ket{ss}$ is suppressed, not because of their interaction in the final state but the excitation proceeds via the state $\frac{1}{\sqrt{2}}[\ket{rs}+\ket{sr}]$, which is detuned by $V_{rs}$. The population of doubly excited Rydberg states thus becomes small $(\sim \Omega^2_M/V^2_{rs})$, and any finite $V_{ss}$ will make the second order process energy non-conserving and further reduce this population. The van der Waals interaction between atoms occupying the rapidly decaying $\ket{r}$
states has less significance for the dynamics. This novel blockade mechanism may have interesting applications for systems consisting of atoms at both close and far mutual distances, such as separate atomic ensembles. We have thus solved the master equation numerically for the case of two interacting Rydberg ensembles, each with a smaller number of atoms ($N=3$)
separated by distances equal to, and greater than, the blockade radius $R^{ss}_6 = (C^{ss}_6/\Omega_R)^{1/6}$ of the van der Waals interaction
between atoms occupying the Rydberg state $\ket{s}$. The atoms within both ensembles are close enough to obey perfect van der Waals Rydberg blockade,
but the inter-ensemble coupling between Rydberg states makes no such assumption, and the effective Hilbert space dimension for the total system is $(2N+1)^2$. In Fig. 4,
the data points show the entangled state fidelity of all atoms obtained at the final time $\tau = 10\mu$s as a function of the RDDI strength
$V_{rs}(R^{ss}_6)$. The left most (blue squares) data points show the results for ensembles separated by $R^{ss}_6$. The van der Waals interaction is not sufficient
to inhibit multiple excitations of the $\ket{s}$ state, but with  $V_{rs}(R^{ss}_6)\ge 30 V_{ss}(R^{ss}_6)$, the RDDI is strong enough to ensure the entangled steady state. The right most (green circles) data points show the fidelity when the two ensembles are separated by a distance of $2R^{ss}_6$,  and in this case a much stronger
RDDI $(V_{rs}(R^{ss}_6)\ge 300 V_{ss}(R^{ss}_6))$ is required to ensure the dark entangled state of all atoms in the two ensembles. Here, transfer of population from $\ket{s}$ to the short-lived $\ket{e}$-state will produce, on demand from the steady state, a single photon or a superposition of zero and one photon in the interference pattern of two sources.


\begin{figure}
\includegraphics[width=80mm]{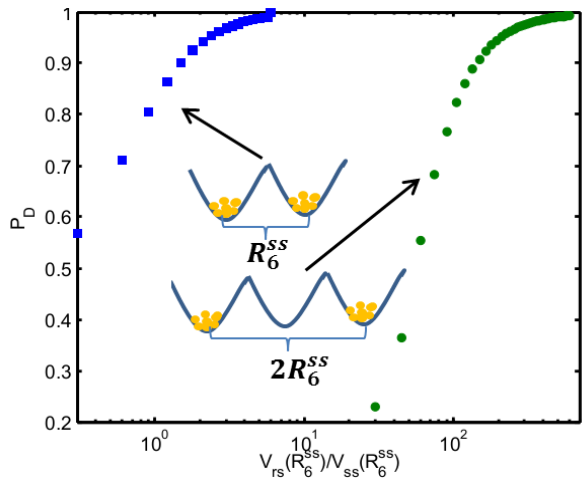}
\label{setup1}
\caption{(Color online) Population of the  dark state $\ket{\psi_D}$ in Eq. 3 with 6 atoms distributed evenly in two remote ensembles. The atoms within each ensemble have an equidistant spacing of $0.1\mu$m and are perfectly blocked while the van der Waals coupling between atoms in different ensembles is too weak to block their double occupancy.
The fidelity is obtained by numerical integration of the master equation until $\tau= 10\mu$s and plotted as a function of the RDDI coupling $V_{rs}$ for ensembles separated by a relative distances $R^{ss}_6 = 3\mu$m and $2R^{ss}_6 = 6\mu$m.}
\end{figure}

In conclusion, we have shown that the interplay of laser excitation of a Rydberg state, microwave driving between Rydberg states and engineered dissipation of one of these states presents a mechanism that drives atomic ensemble towards a dark steady state. The resonant dipole-dipole exchange interaction between atoms populating different Rydberg states plays a crucial role, and it both serves to block the transition path to states with pairs of atom in the same Rydberg state, and to define the unique dark state of the system. With realistic parameters, we have shown that we obtain a good approximation to the W-state, with multiple applications in quantum information science. It should be noted that under the same assumption of strong Rydberg interactions between the Rydberg states it is also possible to apply adiabatic passage and prepare $\ket{\psi^{(N)}_D}$ in Eq.(3) in a unitary manner with time varying fields \cite{lars}. In comparison, dissipative schemes generally benefit from being auto-correcting and robust to dephasing and decay errors \cite{peter, sorensen}. Our entangled steady state may furthermore be obtained, even in the case of a vanishing short range van der Waals interaction between atoms occupying the same Rydberg state, as long as the RDDI with its slower fall-off with distance is sufficiently strong.

Throughout the Letter we have assumed the application of classical laser and microwave fields.
The strong microwave coupling between adjacent Rydberg levels, however, permits strong coupling
to single microwave photons in a high-Q superconducting cavity \cite{haroche} or
coplanar waveguide \cite{davidger, saffsup}. Our results remain valid in this quantized field regime,
where our scheme may be exploited for deterministic storage and retrieval of single
microwave photons.

The authors acknowledge useful discussion with David Petrosyan and Mark Saffman and
financial support from the Villum Foundation and the IARPA MQCO program.

\end{document}